\documentclass[useAMS,usenatbib,usegraphicx]{mn2e}

\title[HD\,145792: a new Helium strong star]
  {Helium stratification in HD\,145792: a new He strong star\thanks{Based on observations
collected at European Southern Observatory (ESO), La Silla, Chile, proposal ID 69.D-0537}}
\author[G. Catanzaro]
  {G.~Catanzaro\thanks{E-mail: Giovanni.Catanzaro@oact.inaf.it}\\
  INAF - Osservatorio Astrofisico di Catania, Via S. Sofia 78, 95123 Catania, Italy}
 \date{Accepted 2007 November 27.  Received 2007 November 14; in original form 2007 October 16}

\pagerange{\pageref{firstpage}--\pageref{lastpage}} \pubyear{2007}

\def\LaTeX{L\kern-.36em\raise.3ex\hbox{a}\kern-.15em
    T\kern-.1667em\lower.7ex\hbox{E}\kern-.125emX}

\begin{document}

\label{firstpage}

\maketitle

\begin{abstract}
In this paper we report on the real nature of the star HD\,145792, classified as He weak
in {\it ``The General Catalogue of Ap and Am stars''}. By means of FEROS@ESO1.52m 
high resolution spectroscopic data, we refined the atmospheric
parameters of the star, obtaining: T$_{\rm eff}$\,=\,14400~$\pm$~400~K, $\log g$\,=\,4.06~$\pm$~0.08
and $\xi$\,=\,0 $^{+0.6}$ km s$^{-1}$. These values resulted always lower than those derived by 
different authors with pure photometric approaches.

Using our values we undertook an abundance analysis with the aim to derive, for the first time,
the chemical pattern of the star's atmosphere. For metals a pure LTE synthesis (ATLAS9 and SYNTHE)
has been used, while for helium a hybrid approach has been preferred (ATLAS9 and SYNSPEC).
The principal result of our study is that HD\,145792 belongs to He strong class contrary to the
previous classification. Moreover, helium seems to be vertically stratified in the atmosphere,
decreasing toward deepest layers. 

For what that concerns metals abundances, we found the following: overabundance of oxygen, 
neon, silicon, phosphorus, sulfur and calcium; carbon, nitrogen, magnesium, aluminum,
titanium, chromium and nickel are normal, being the discrepancies from the solar values 
within the experimental errors; iron resulted to be slightly underabundant.

\end{abstract}


\section{Introduction}
HD\,145792 (=HR\,6042\,=V\,1051~Sco) is a chemically peculiar star belonging to the
Sco-Cen OB association (\citet{morris61}; \citet{eggen98}) and classified as Helium weak 
star in the {\it General Catalogue of Ap and Am stars} by \citet{renson91}.
Hipparcos photometry has been analyzed by \citet{paunzen98}, they discovered light variability
with a period of 0.84780 days. A few magnetic field measurements are present in the 
catalogue by \citet{bychkov03} with an average value of H$_{\rm eff}$\,=\,285~$\pm$~190 gauss.

Regarding atmospheric parameters, there are in the literature several inconsistent determinations 
of $T_{\rm eff}$ and $\log g$. For instance, \citet{nissen74} found 
$T_{\rm eff}$\,=\,16000~K and $\log g$\,=\,4.34, 
\citet{geus89} found $T_{\rm eff}$\,=\,15135~K and $\log g$\,=\,4.38 and \citet{gulati89}
found effective temperatures ranging between 18450~K and 18920~K. All these
studies are based on photometric approach. No previous abundances determination have been
reported in the literature.

The principal aim of this study was to determine for the first time the chemical abundances
in the HD\,145792's atmosphere, with particular attention to helium behavior. With this goal 
in mind, we obtained high resolution spectrum of the star with FEROS@ESO1.52~m telescope. 
Atmospheric parameters like $T_{\rm eff}$, $\log g$\ and microturbulence have been derived 
by us with spectroscopic methods. In the following sections we present our results, giving 
particular emphasis to the actual peculiarity of the star: according to our results HD\,145792 
is a Helium strong star contrarily to the previous classification reported in \citet{renson91}.
  
\begin{figure*}
\includegraphics[width=12cm]{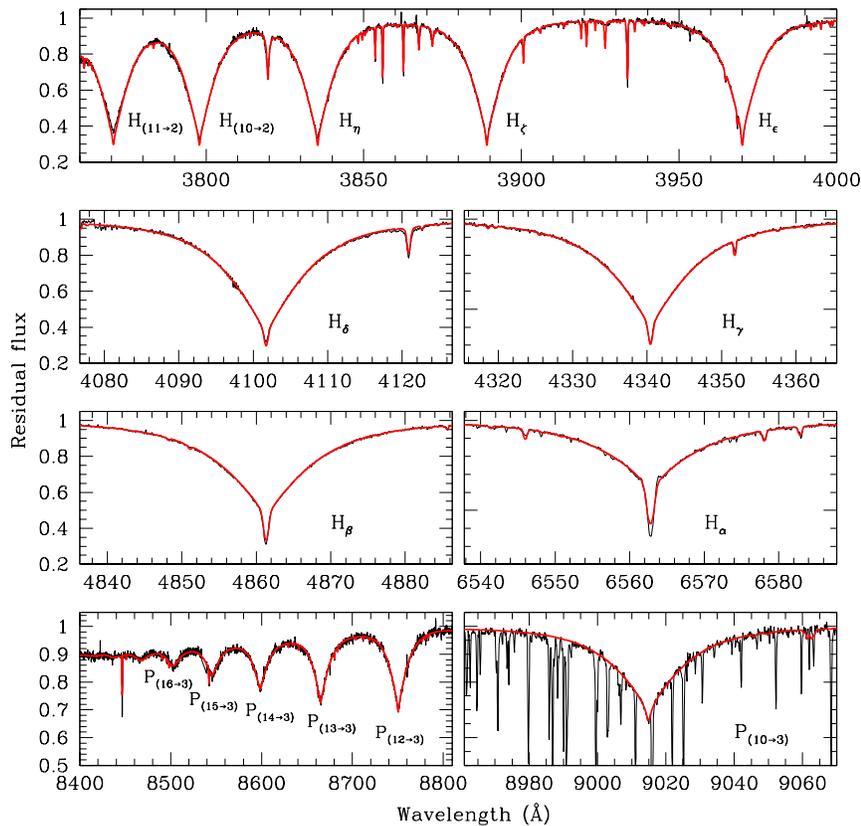}
\caption[]{Comparison between the observed and computed hydrogen line
  profiles. From top to bottom: Balmer series from H$_{(n=11 \rightarrow 2)}$
to H$_{\alpha}$ and Paschen series from P$_{(n=17 \rightarrow 3)}$ to
P$_{(n=10 \rightarrow 3)}$. The readers should note the strong telluric lines 
contamination around this line.}
\label{balmer}
\end{figure*}

\section{Observation and data reduction}
A spectrum of HD\,145792 in the 3600 - 9200 {\AA} spectral range was obtained with the FEROS 
spectrograph on August 30, 2002 (HJD\,=\,2\,452\,516.509) at the ESO 1.52\,m telescope at 
La Silla Observatory, Chile. The spectral resolution was R\,=\,48000. 

The stellar spectrum, wavelength calibrated and normalized to the continuum, was obtained 
using standard data reduction procedures for spectroscopic observations within the NOAO/IRAF 
package. The resulting signal-to-noise ratio was $\approx$~200.

\section{Atmospheric parameters}
\label{atmos}
The approach we used in this paper was to minimize the difference between observed and
synthetic H$_{\delta}$, H$_{\gamma}$ and H$_{\beta}$ profiles. As goodness-of-fit test we used the 
parameter:
\\

$\chi^2 = \frac{1}{N} \sum (\frac{I_{\rm obs} - I_{\rm th}}{\delta I_{\rm obs}})^2$

\bigskip

\noindent
where $N$ is the total number of points, $I_{\rm obs}$ and $I_{\rm th}$ are the intensities
of the observed and computed profiles, respectively, and $\delta I_{\rm obs}$ is the photon noise.
The synthetic spectra were generated in three  steps: first, we computed the stellar 
atmosphere model by using the ATLAS9 code \citep{kur93} then, the stellar spectrum 
was synthesized using SYNTHE \citep{kur81} and finally, the instrumental and 
rotational convolutions were applied. The ATLAS9 code includes the metal opacity by means of
distribution functions (ODF) that are tabulated for multiples of the solar metalicity
and for various microturbulence velocities. 

\citet{leo97a} determined $T_{\rm eff}$, $\log g$, microturbulence 
velocity and abundances, minimizing $\chi^2$ adopting an iterative procedure, but for H$_\beta$
alone. Later, \citet{catanzaro04} extended this method using H$_{\delta}$ and
H$_{\gamma}$ as $T_{\rm eff}$ and $\log g$ indicators for non-solar
composition atmospheres. In this study we extend the previous method to three Balmer
lines: H$_{\delta}$, H$_{\gamma}$ and H$_\beta$. The intersection
of the three $\chi^2$ iso-surfaces is expected to improve the final
solution. Besides H$_{\alpha}$ is present in our spectral range, we did not use it principally 
for two reasons: {\it a)} it is located too close to the edge of the echelle order to be
correctly normalized and {\it b)} for the intrinsic difficulties in modeling the
line core with pure LTE approach.

To decrease the number of parameters, first of all we computed the $v_e \sin i$ of our star.
To derive the rotational velocity, we used SYNTHE to reproduce the profile of
Mg{\sc ii}~$\lambda$4481 {\AA} line.
Contrary to the value of 30 km s$^{-1}$ reported by \citet{abt02}, our 
best match with the observations was achieved convolving the computed profiles with a
stellar rotational profile having $v_e \sin i$\,=\,18 km s$^{-1}$. As a by-product
of this calculation we derived also its radial velocity: RV\,=\,$-$4.34~$\pm$~0.43 km s$^{-1}$.

Then, to determine stellar parameters as consistent as possible with
the actual structure of the atmosphere, we have reproduced the Balmer lines by
the following iterative procedure:

\begin{figure}
\includegraphics[width=9cm,height=10cm]{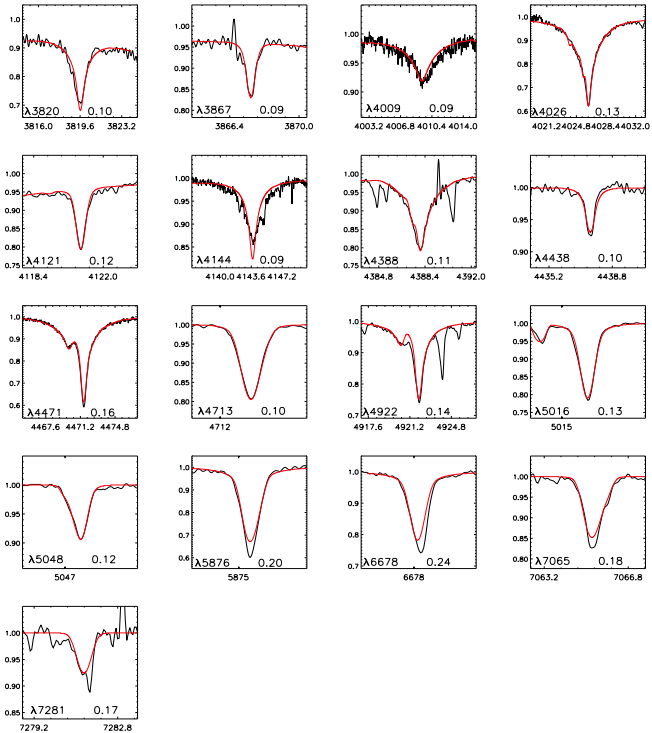}
\caption[]{Comparison between observed and synthetic He I lines profiles. Due to the 
contamination of telluric lines the fits are not satisfactory for the last four lines. 
In the other cases only helium lines have been considered, that is the reason of absence 
of metallic lines. In each box we reported also the theoretical wavelength and the inferred
abundance expressed as n(He)/n(H).}
\label{helium}
\end{figure}

\begin{itemize}

\item a) as starting values of T$_{\rm eff}$ and $\log g$ we have determined the 
effective temperature and gravity from Str\"omgren photometry according to 
the grid of \citet{md85}. The photometric colors have been 
de-reddened with the \citet{moon85} algorithm. The source of the 
Str\"omgren photometric data was \citet{hauck98}.
Results of our calculations are: T$_{\rm eff}$\,=\,15320\,$\pm$\,50~K and 
$\log g$\,=\,4.27\,$\pm$\,0.04;

\item b) determination of $T_{\rm eff}$ and $\log g$ of the ATLAS9 atmosphere model whose 
H$_{\delta}$, H$_{\gamma}$ and H$_{\beta}$ profiles, computed with SYNTHE, match the observations. 
The model has been computed using the opacity scale and abundances of the Sun;

\item c) helium and metals abundances are derived and the microturbulent velocity
determined independently from two sets of 10 Si{\sc ii} and 15 Fe{\sc ii} lines\footnote{For the
purpose of $\xi$ determination we used all the lines with EW $>$ 10 m{\AA}} 
by requiring that the derived abundances do not depend on the measured equivalent widths;

\item d) thus, we repeated the calculations described in step b) with the correct helium abundance,
solar metalicity and $\xi$. For model calculations, we used a step of 50~K for T$_{\rm eff}$ 
and 0.01 dex for $\log g$. 

\end{itemize}

The best fit was obtained for the model computed with the ODF for [M/H]=0.0, solar helium and:
\\
\\
\noindent
T$_{\rm eff}$\,=\,14400 $\pm$ 400 K\\
$\log g$\,=\,4.06 $\pm$ 0.08 \\
$\xi$\,=\,0 $^{+0.6}$ km s$^{-1}$\\

Errors on T$_{\rm eff}$ and $\log g$ have been estimated within 1~$\sigma$ level of
confidence, as the variation in the parameters which increases the $\chi^2$ of a
unit \citep{lampton76}.

The synthetic Balmer lines compared with the observed ones are showed in
Fig.~\ref{balmer}. 
With the atmospheric parameters found as described before, we attempted also
in modeling the other hydrogen lines present in our spectral range, in
particular from H$_{\epsilon}$ to the limit of the Balmer series (top panel
of Fig.~\ref{balmer}), H$_\alpha$ and the Paschen series from its limit at
$\lambda$ 8400 {\AA} to the P$_{(n\,=\,10 \rightarrow 3)}$ at $\lambda$ 9012
{\AA} (bottom panels of Fig.~\ref{balmer}). The good modeling of Paschen series
with LTE synthesis, should not be a surprise as shown by \citet{przybilla04}.

\begin{table}
\caption[]{Summary of the inferred abundances for each neutral helium lines detected in our 
spctrum. The three red lines without errors are those strongly contaminated by telluric 
contributions.}
\label{ab_helium}
\begin{center}
\begin{tabular} {cl}
\hline
\hline
$\lambda$ ({\AA}) & ~n(He)/n(H)  \\
\hline
3820.600 & 0.10~$\pm$~0.01  \\
3867.483 & 0.09~$\pm$~0.02  \\
4009.257 & 0.09~$\pm$~0.02  \\
4026.210 & 0.13~$\pm$~0.03  \\
4120.811 & 0.12~$\pm$~0.03  \\
4143.759 & 0.09~$\pm$~0.01  \\
4387.929 & 0.11~$\pm$~0.02  \\
4437.500 & 0.10~$\pm$~0.02  \\
4471.498 & 0.16~$\pm$~0.03  \\
4713.139 & 0.10~$\pm$~0.03  \\
4921.931 & 0.14~$\pm$~0.03  \\
5015.678 & 0.13~$\pm$~0.02  \\
5047.738 & 0.12~$\pm$~0.03  \\
5875.614 & 0.20             \\
6678.152 & 0.24             \\
7065.119 & 0.18             \\
7281.349 & 0.17~$\pm$~0.05  \\
\hline
\end{tabular}
\end{center}
\end{table}

\section{Abundance analysis}
Below we describe and comment on the abundance analysis for helium and metals in turn:

\subsection{Helium} 
Neutral helium lines in a normal star with T$_{\rm eff}~\approx$\,13000 
are usually strong, even if the maximum strength is reached for the spectral
type B2. To investigate the helium abundance 
we attempted to reproduce, by spectral synthesis, all the lines profiles
observed in our spectrum.

Non-LTE effects play an important role in the formation of helium line, as known
since the pioneering work by \citet{auer72}. Thus, because of the importance of 
NLTE effects, the synthetic helium lines were computed with the version 43 of the 
program SYNSPEC \citep{hubeny00}.
This program reads the same input model atmosphere previously computed using ATLAS9
and solves the radiative transfer equation, wavelength by wavelength in a specified 
spectral range. SYNSPEC also reads the same Kurucz list of lines we used for the metal 
abundances. SYNSPEC allows one to compute the line profiles considering an approximate
NLTE treatment even for LTE models. This is done by means of the second-order escape
probability theory (for details see the paper by \citet{hubeny86}).
Moreover, the helium atom is considered explicitly and 14 levels of He{\sc i} are taken 
into account. For  He{\sc i} $\lambda$4471 {\AA} detailed line broadening tables are 
taken from \citet{barnard74}, for $\lambda \lambda$4026, 4387 and 4922 {\AA} from
\citet{shamey69} and for the other lines from \citet{dimi84}.
Results of our modeling are displayed in Fig.~\ref{helium}, while inferred abundances
with their errors have been reported in Tab.~\ref{ab_helium}. These errors have been
derived translating errors on atmospheric parameters into abundances uncertainties. 
Because of the presence of telluric lines contamination the fit of the lines:
He{\sc i} $\lambda \lambda$5875, 6678 and 7065 {\AA}, is not satisfactory and the 
derived abundances have to be considered as low limits.

The first important result of this analysis is that helium is not underabundant, then 
HD\,145792 could be ruled out from the He weak sub-class.

According to the calculations by \citet{vauclair91}, helium is stratified in the atmospheres
of magnetic CP stars: helium abundance increases with optical depth, reaches a maximum and 
then decreases. The position of the helium abundance maximum depends essentially on
effective temperature, mass loss and diffusion strength. Vauclair's calculations have been 
verified for He strong stars by \citet{leo97b} and \citet{leo98}. These authors performed 
spectroscopic observations in a sample of helium strong stars obtaining that helium abundance
decrease going down through the atmospheres.

\begin{figure}
\includegraphics[width=9cm]{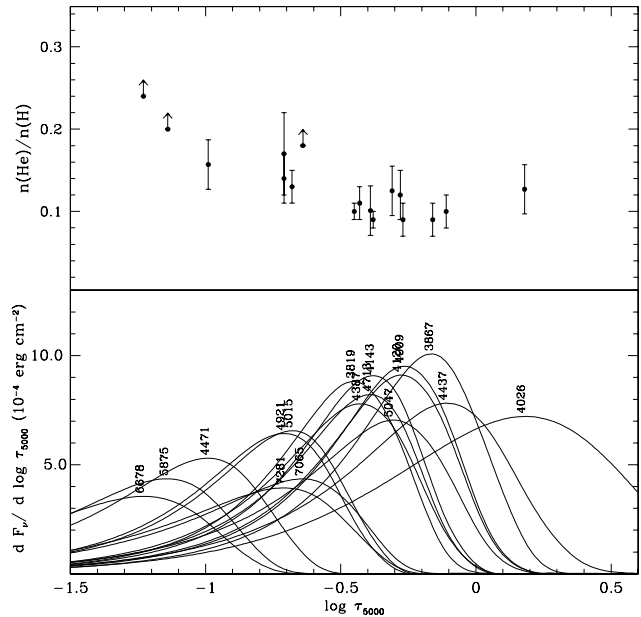}
\caption[]{Helium stratification measured in HD\,145792. The bottom panel shows the contribution
functions computed for the lines core.}
\label{helium_str}
\end{figure}

In order to search for helium stratification in HD\,145792,
we computed the contribution functions for the line cores of all helium lines reported 
in Fig.~\ref{helium}, calculations have been performed using the code XLINOP \citep{kur81}.
In the upper panel of Fig.~\ref{helium_str} we reported our abundances as a function of
optical depth. We observed a constant and almost solar helium abundance 
(n(He)/n(H)~$\approx$~0.1) in the range of optical depths between 0.2 and -0.5, then helium 
seems to increase its abundance (up to n(He)/n(H)~$\approx$~0.15) going toward external layers.
From this conclusion, we have excluded the three lines for which the strong telluric contamination
did not allow us a good fit. Actually, if their contribute should be considered, helium
abundance will increase up to n(He)/n(H)~$\approx$~0.25. In this case the effects of the
change of the mean molecular weight in the atmospheric structure should be included. 

Another important mark of the presence of stratification in the impossibility to reproduce
the profile of a strong line with one abundance only. If we look at the $\lambda$4026 {\AA}
we note that while the core is well reproduced the red wing is not. In particular the helium
line at $\lambda$4023.973 {\AA} computed with the same abundance is stronger than that observed. 
This could be interpreted as another evidence in favor of stratification hypothesis.

\subsection{Metals}
To derive the metals abundances we identified all the unblended lines 
in the observed spectrum of HD\,145792 using SYNTHE. The atomic parameters adopted in our 
analysis are from \citet{kur95} line lists and subsequent update by \citet{castelli04}. 
In practice we divided all the spectrum in sub-intervals $\approx$100 {\AA} wide and for each 
interval we performed a synthesis analysis. The final adopted abundances are the weighted
averages of these values and they are expressed in the usual form $\log N_{el}/N_{Tot}$.

\begin{figure}
\includegraphics[width=9cm]{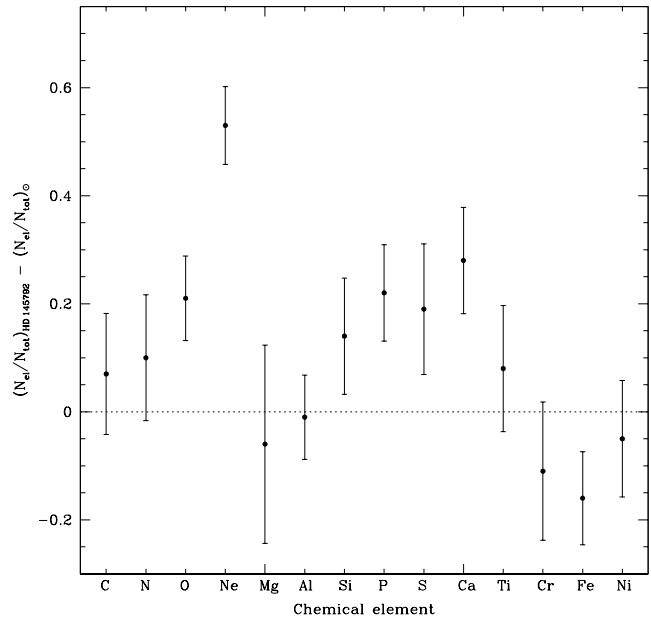}
\caption[]{Abundance pattern computed for all metals detected in our spectrum.}
\label{patt}
\end{figure}

Derived abundances are reported in Tab.~\ref{abundances} with their errors, in the
same table for the sake of comparison we reported also the Sun abundances taken from
\citet{asplund05}. Errors on abundances have been estimated as the following:
for all the elements for which we observed more than 5 spectral lines, we simply calculated
the standard deviation; for the other cases we translated the errors on T$_{\rm eff}$ and
$\log g$ in abundances uncertainties. The chemical pattern of HD\,145792 is shown in Fig.~\ref{patt}. 
The principal result is an evident overabundance of neon with respect the solar case, a slight 
overabundance of oxygen, silicon, phosphorus, sulfur and calcium, iron appears to be slightly 
under the solar value and other elements are normal, being their discrepancies from the solar 
values within the experimental errors. 

\begin{table}
\caption[]{Summary of the inferred abundances in the atmosphere of HD\,145792. In the second
column we reported the number of spectral lines used for the abundance determination. For comparison
we reported the relative abundance in the solar photosphere \citep{asplund05}.}
\label{abundances}
\begin{center}
\begin{tabular} {lrcc}
\hline
\hline
El & N & \multicolumn{2}{c}{$\log N_{el}/N_{Tot}$}   \\
   &   & ~~HD\,145792 & Sun \\
\hline
C  &  5 & $-$3.57 $\pm$ 0.10 & $-$3.64 $\pm$ 0.05\\
N  &  5 & $-$4.15 $\pm$ 0.10 & $-$4.25 $\pm$ 0.06\\
O  & 15 & $-$3.16 $\pm$ 0.06 & $-$3.37 $\pm$ 0.05\\
Ne & 17 & $-$3.66 $\pm$ 0.04 & $-$4.19 $\pm$ 0.06\\
Mg &  7 & $-$4.56 $\pm$ 0.16 & $-$4.50 $\pm$ 0.09\\
Al &  9 & $-$5.67 $\pm$ 0.05 & $-$5.66 $\pm$ 0.06\\
Si & 19 & $-$4.38 $\pm$ 0.10 & $-$4.52 $\pm$ 0.04\\
P  &  2 & $-$6.45 $\pm$ 0.08 & $-$6.67 $\pm$ 0.04\\
S  & 32 & $-$4.70 $\pm$ 0.11 & $-$4.89 $\pm$ 0.05\\
Ca &  3 & $-$5.44 $\pm$ 0.09 & $-$5.72 $\pm$ 0.04\\
Ti &  2 & $-$7.05 $\pm$ 0.10 & $-$7.13 $\pm$ 0.06\\
Cr &  3 & $-$6.50 $\pm$ 0.08 & $-$6.39 $\pm$ 0.10\\
Fe & 67 & $-$4.74 $\pm$ 0.07 & $-$4.58 $\pm$ 0.05\\
Ni &  1 & $-$5.85 $\pm$ 0.10 & $-$5.80 $\pm$ 0.04\\
\hline
\end{tabular}
\end{center}
\end{table}

\section{Conclusion}
In this paper we report on the first abundances analysis ever presented in the literature
on the star HD\,145792. Principal results of our paper can be summarized in this
two important points:

\begin{itemize}

\item the object resulted to be a He strong star, contrarily to the classification reported
in \citet{renson91}. The erroneous classification could be linked to effective temperatures
and gravities reported in the literature that, as we discussed in Sect.~\ref{atmos}, are always 
greater than ours ($T_{\rm eff}$\,=\,14400~$\pm$~400~K and $\log g$\,=\,4.06~$\pm$~0.08).

\item helium seems to be not homogeneously distributed along the atmosphere. It shows 
approximatively a constant and almost solar abundance from the inner layers up to 
$\log \tau_{5000}$\,=\,-0.5 and a moderate increase toward the outer atmospheric layers 
(see Fig.~\ref{helium_str})

\end{itemize}

For what that concern the chemical pattern of metals, we found, with respect the solar case,  
a strong overabundance of neon; a slight overabundance of oxygen, silicon, 
phosphorus, sulfur and calcium; iron appears to be slightly underabundant; the others
elements are normal.

\section*{Acknowledgments}
This research has made use of the SIMBAD database,
operated at CDS, Strasbourg, France

\end{document}